\documentclass{ws-procs975x65}

\begin{document}

\wstoc{The final fate of binary neutron stars}{M. Duez}

\title{THE FINAL FATE OF BINARY NEUTRON STARS: \\
  WHAT HAPPENS AFTER THE MERGER?}

\author{MATTHEW D. DUEZ}

\address{Department of Astronomy, Cornell University, Ithaca, NY 14850 USA\\
\email{mduez@astro.cornell.edu}}

\author{YUK TUNG LIU, STUART L. SHAPIRO, BRANSON STEPHENS}

\address{Department of Physics, University of Illinois at Urbana-Champaign,
  Urbana, IL 61801-3080 USA}

\author{MASARU SHIBATA}

\address{Graduate School of Arts and Sciences, University of Tokyo, Komaba, Meguro, Tokyo 153-8902, Japan}

\begin{abstract}
The merger of two neutron stars usually produces a remnant with a mass
significantly above the single (nonrotating) neutron star maximum mass. 
In some cases, the remnant will be stabilized against collapse by rapid,
differential rotation.  MHD-driven angular momentum transport eventually
leads to the collapse of the remnant's core, resulting in a black hole
surrounded by a massive accretion torus.  Here we present
simulations of this process.  The
plausibility of generating short duration gamma ray bursts through this
scenario is discussed.
\end{abstract}

\bodymatter

\section{Introduction}\label{intro}
The merger of binary neutron stars is now one of
the favored hypotheses for explaining short gamma ray bursts (GRBs).
According to this 
scenario, after the merger, a stellar-mass black hole (BH) is formed with an ambient 
accretion torus with $\sim 1$--$10\%$ of the total mass.  Energy extracted 
from this system by MHD processes or neutrino radiation powers 
the GRB fireball.  The viability of this model depends on the presence 
of a significantly massive accretion disk after the collapse of the
remnant 
following merger, and the presence of a baryon-poor region above the accretion disk. 

A typical binary neutron star system has total mass
2.6--$2.8M_{\odot}$~\cite{Stairs}, much larger than the spherical neutron
star maximum mass $M_{\rm TOV}$.  However, the remnant is also
rapidly and differentially rotating.  Mass limits for nonrotating stars
and for rigidly rotating stars (the {\it supramassive} limit,
$M_{\rm sup} \approx 1.2 M_{\rm TOV}$) can be significantly exceeded
when differential rotation is present~\cite{BSS}.  Stars with
masses greater than $M_{\rm sup}$ are called {\it hypermassive} stars.
Thus, the remnant may form a hypermassive neutron star (HMNS).
General relativistic hydrodynamic simulations
have shown that just after the merger,
either a black hole or a HMNS is formed \cite{STU2}.  A BH forms
promptly if the total mass of the system, $M$, is larger than a
threshold mass $M_{\rm thr}\approx 2.8 M_{\odot}$.  In this case, far less
than 1\% of the matter remains outside the horizon, which is unfavorable
for GRBs.  On the other hand, for $M < M_{\rm thr}$, a HMNS forms.

These HMNSs may survive for many orbital
periods.  However, on longer timescales magnetic fields will transport
angular momentum and may trigger gravitational collapse.
Two important mechanisms which transport angular momentum are magnetic
braking~\cite{BSS,Shapiro} and the magnetorotational instability
(MRI)~\cite{MRI}.  Magnetic breaking
transports angular momentum on the Alfv\'en time
scale~\cite{BSS,Shapiro}, $\tau_A \sim R/v_A \sim 10^2 (B/10^{12}~{\rm
G})^{-1}~{\rm s}$, 
where $R$ is the radius of the HMNS. MRI occurs wherever angular
velocity $\Omega$ decreases with cylindrical radius $\varpi$.  This
instability grows exponentially with an e-folding time of $\tau_{\rm
MRI} =4 \left(\partial \Omega/\partial \ln \varpi
\right)^{-1}$~\cite{MRI}, independent of the field strength. For the
HMNS model considered here, $\tau_{\rm MRI} \sim 1$~ms.  The length scale
of the fastest growing unstable MRI modes,
$\lambda_{\rm MRI}$, does depend on the field strength: 
$\lambda_{\rm MRI}\sim 3~{\rm cm}$ $(\Omega/4000 s^{-1})^{-1}$
$(B/10^{12}{\rm G}) \ll R$. 
When the MRI saturates, turbulence consisting of small-scale eddies
often develops, leading to angular momentum transport on a
timescale much longer than $\tau_{\rm MRI}$ \cite{MRI}. 

\section{Simulations}\label{sims}
To determine the final fate of the HMNS, it is necessary to carry out
magnetohydrodynamic simulations in
full general relativity.  Such simulations have only recently become
possible.  Duez {\it et al.}~\cite{DLSS2} and Shibata and Sekiguchi~\cite{SS}
have  developed new codes to evolve the full Einstein-Maxwell-MHD system
of equations self-consistently.  These codes have since been used to simulate
the evolution of magnetized hypermassive neutron stars~\cite{DLSSS1,bigpaper},
and
implications for short GRBs have been investigated~\cite{GRB2}.

We assume axisymmetry and equatorial symmetry in all our simulations. 
We use uniform computational grids with sizes
up to 500$\times$500.  To model the
remnant formed in binary merger simulations, we use as our initial data an
equilibrium HMNS, a $\Gamma=2$ polytrope, with mass
$M = 1.7 M_{\rm TOV} = 1.5 M_{\rm sup}$ and a rotation profile
chosen so that
the ratio of equatorial to central $\Omega$ is $\sim 1/3$. 
(We find that an HMNS with a more realistic equation of state evolves
similarly~\cite{GRB2,bigpaper}.)  We add a
poloidal magnetic field with strength 
proportional to the gas pressure.  The initial magnetic pressure
is set much smaller than the gas pressure, but not so small that
$\lambda_{\rm MRI}$ cannot be resolved.  Therefore, we set
$\lambda_{\rm MRI} \approx R/10$, corresponding to $B \approx 10^{16}$ G
and max$(B^2/P) \sim 10^{-3}$.

In our evolutions, the effects of magnetic winding are observed in the
generation of a toroidal $B$ field which grows linearly with time 
during the early phase of the evolution, and saturates on the Alfv\'en 
timescale.  The
effects of MRI are observed in an exponential growth of the poloidal field
on the $\lambda_{\rm MRI}$ scale, a growth which saturates after a few
rotation periods.  The magnetic fields cause angular momentum to be
transported outward, so that the core of the star contracts while the outer
layers expand.  After about 66 rotation periods, the core collapses to a
black hole.  Using singularity excision~\cite{excision}, we continue
the evolution to a quasi-stationary state.  The final state consists of a
black hole of irreducible mass $0.9M$ surrounded by a hot accreting torus
with rest mass $0.1M$ and a collimated magnetic field near the polar region. 
At its final accretion rate, the torus should survive $\sim$10ms.  The
torus is optically thick to neutrinos, and we estimate that it will emit
$\sim 10^{50}$ergs in neutrinos before being accreted.  We also find that the region
above the black hole is very baryon-poor. All these properties make this system a promising central
engine for a short-hard GRB.

In order to study the evolution of the magnetic field more realistically,
it will be necessary to redo the evolutions in three dimensions.  Also,
realistic
merger remnants are usually expected to have much smaller magnetic fields
than the ones studied here.  In an earlier analysis, we modeled the effects
of small-scale MRI turbulence as a shear viscosity~\cite{visc} and found that,
if this model is valid, the evolution of the HMNS is qualitatively similar
to that shown here.

\section{Acknowledgments}
Numerical computations were performed at the National Center for
Supercomputing Applications at the University
of Illinois at Urbana-Champaign (UIUC), and
on the FACOM xVPP5000 machine at
the data analysis center of NAOJ and the NEC SX6 machine in ISAS,
JAXA. This work was in part supported by NSF Grants PHY-0205155
and PHY-0345151, NASA Grants NNG04GK54G and NNG046N90H
at UIUC, and
Japanese Monbukagakusho Grants (Nos.\ 17030004 and 17540232).

\vfill

\end{document}